\pdfoutput=1
\documentclass[10pt, conference]{IEEEtran}

\usepackage{cite}
\usepackage[latin1]{inputenc}
\usepackage[cmex10]{amsmath}
\usepackage{amsmath}
\usepackage{amsfonts}
\usepackage{amssymb}
\usepackage{graphicx}
\usepackage{graphics}
\usepackage{lscape}
\usepackage{comment}
\usepackage{epstopdf}
\usepackage{array}
\usepackage{color}

\numberwithin{theorem}{section}
\newtheorem{lemma}{Lemma}
\numberwithin{lemma}{section}
\newtheorem{prop}{Proposition}
\numberwithin{prop}{section}
\newtheorem{corol}{Corollary}
\numberwithin{corol}{section}

\numberwithin{remark}{section}
\newtheorem{conject}{Conjecture}
\numberwithin{conject}{section}

\begin{document}

\title{User Performance in Small Cells Networks 
\\
with Inter-Cell Mobility}


\author{\IEEEauthorblockN{P. Olivier, A. Simonian}
\IEEEauthorblockA{Orange Labs, Ch\^atillon, France
\\
E-mail: \{alain.simonian, phil.olivier\}@orange.com}}

\maketitle

\begin{abstract}
We analyze the impact of inter-cell mobility on user performance in dense networks such as that enabled by LTE-A and 5G. To this end, we consider a  homogeneous network of small cells and first show how to reduce the evaluation of user performance to the case of a single representative cell. We then propose simple analytical models that capture mobility through the distribution of the residual sojourn time of mobile users in the cell. An approximate model, based on Quasi-Stationary (QS) assumptions, is developed in order to speed up computation in the Markovian framework. We use these models to derive the average throughput of both mobile and static users, along with the probability of handover for mobile users. Numerical evaluation and simulation results are provided to assess the accuracy of the proposed models. We show, in particular, that both classes of users benefit from a throughput gain induced by the ``opportunistic'' displacement of mobile users among cells. 
\end{abstract}



\section{Introduction}


To address the steady increase of mobile traffic, one way is to enhance the network capacity by a massive deployment of small cells. This is a solution envisaged by network operators in the framework of LTE-A heterogeneous networks \cite{Dah13} or Ultra Dense Networks scenarios for future 5G networks \cite{Oss14}.  In such dense networks, the amount of handover generated by users mobility will significantly increase, with a notable impact on signaling overhead. On the other hand, the average throughput of users data communications is expected to improve, due to the changing transmission conditions which allow some kind of mobility-based opportunism. 

In this context, the present paper aims at evaluating the impact of \textit{inter-cell mobility} on the user-perceived performance in dense networks. Specifically, considering small cells enables us to neglect the possible spatial variations of intra-cell capacities and thus to focus on the impact of inter-cell mobility itself. Furthermore, we decouple the performance evaluation problem from that of characterizing a geometric user displacement model. The latter topic is out of scope of the present paper (see \cite{Schin06} and references therein for a review of current methods and, in particular, \cite{Xing13} for the Random Waypoint (RW) model). Mobility is here supposed to be captured through the distribution of the users residual sojourn time in a cell, i.e., the time a mobile user is physically present in the cell once its transmission has started; this distribution may be the output of user displacement models. Given this distribution, we construct a flow-level queuing model that allows us to derive the essential performance metrics in each cell, namely the mean throughput and the handover probability. In the case of a homogeneous network, our approach enables us to reduce the analysis of the network performance to that of a single representative cell. Besides, an approximate model is derived in order to replace the exact Markovian model when the latter suffers from scalability impairments.

In the literature, performance gains induced by mobility in cellular networks are generally related to the spatial variations of capacity inside the cells, which permits some opportunistic use of favorable transmission conditions by mobile users \cite{Bon04, Bon09, Bor06, Kar11}. These papers also base their evaluation on flow-level traffic modeling but address mobility through a spatial Markov process where users jump between distinct capacity zones in the cells. In \cite{Bon04, Bon09}, the average throughputs of different classes of users are estimated through upper and lower bounds associated with fluid limits and quasi-stationary regimes, respectively. By means of an approximate model, \cite{Kar11} computes the mean flow throughput of a single class of users, reporting on a throughput gain created by intra- and inter-cell mobility altogether. In a complementary way, \cite{Bor06} focuses on the characterization of network capacity with different scenarios of intra-cell and inter-cell mobility. 

Generalizing a study on a single cell in isolation \cite{Bay15}, our paper is to our knowledge the first attempt to focus on the multi-class throughput gains induced by inter-cell mobility alone, and also to provide a method to estimate the handover rate, a key performance parameter which counterbalances the throughput improvement. In the following, Section 
\ref{Sec:Models} presents our approach to model networks with mobility. A generic one-cell Markovian model is constructed in Section \ref{Sec:MarkovMod} to derive the main performance indicators; a heuristic Quasi-Stationary approximation is then proposed to alleviate computation. Finally, the approach is validated for homogeneous networks through simulation and numerical experiments in Sections \ref{Sec:NumRes}-\ref{Sec:Discuss}.

\section{Network with mobility}
\label{Sec:Models}

Consider a cellular network of $I$ cells with possibly distinct capacities.  Users from $K$ traffic classes may appear and move during their communications. 
When leaving a cell during transmission, users join one of the neighboring cells according to some routing probabilities. They consequently generate supplementary flows of new arrivals, hereafter called 
\textit{handover arrivals}, which are to be added to \textit{fresh arrivals} in each cell.

We assume that class-$k$ users generate requests for transmission in cell 
$i$ according to a Poisson process with rate $\lambda_{i,k}^0$, 
$i = 1,...,I$, $k=1,...,K$; this corresponds to the fresh traffic offered to cell $i$. To account for class-$k$ users that became active outside cell $i$ and experienced one or more handovers, the total flow arrival to cell $i$ is written as
\begin{equation}
\lambda_{i,k} = \lambda_{i,k}^0 + \lambda_{i,k}^{In}
\label{flow0}
\end{equation}
where $\lambda_{i,k}^{In}$ denotes the handover call arrival rate from neighboring cells. For all $i$ and $k$, we will assume that the handover process to cell $i$ from class-$k$ users can be approximated by a Poisson process so that it can be superposed to the fresh arrivals to build up a total Poisson arrival process with rate $\lambda_{i,k}$ given in (\ref{flow0}). All Poisson processes introduced above are assumed to be mutually independent.

The latter is a fundamental assumption in our modeling approach, which allows a local description of the traffic flows at each cell as explained below. This notably simplifies the global description of the overall multi-class multi-cell process, as handled in \cite{Bon04, Bon09, Bor06, Kar11}.

\label{Sec:FXPT}

\subsection{Heterogeneous network}
\label{Sec:HeterCells}

The handover arrival rate $\lambda_{i,k}^{In}$ to cell $i$ results from the superposition of handovers departing from neighboring cells according to routing probabilities $p_k(j,i)$ from cell $j \neq i$ to cell $i$ for 
class-$k$ users. We can write the flow equation 
\begin{equation}
\lambda_{i,k}^{In} = \sum_{j \neq i} \, p_k(j,i) \cdot \lambda_{j,k}^{Out}
\label{flow1}
\end{equation}
where $\lambda_{j,k}^{Out}$ is the handover departure rate from cell $j$ emanating from class $k$-users. Rate $\lambda_{j,k}^{Out}$ can in turn be considered as an output of a queuing model for cell $j$ and calculated by means of some \textit{performance function} $\mathcal{F}_{j,k}(.)$, that is,
\begin{equation}
\lambda_{j,k}^{Out} = 
\mathcal{F}_{j,k} \left( \lambda_{j,1}^{In}, ..., \lambda_{j,K}^{In} \right )
\label{flow2}
\end{equation}
for $j=1,...,I$ and $k=1,...,K$. In (\ref{flow2}), only handover arrival rates are considered as variables, all other intrinsic parameters (such as cell capacities, per-class offered traffic and mobility rates introduced 
below) being kept constant. From (\ref{flow1})-(\ref{flow2}), it follows that a stationary network regime can be characterized by a system of $I \times K$ equilibrium equations with  unknowns, the handover arrival rates $\lambda_{i,k}^{In}$, namely
\begin{equation}
\lambda_{i,k}^{In} = \sum_{j \neq i} \, p_k(j,i) \cdot 
\mathcal{F}_{j,k} \left( \lambda_{j,1}^{In}, ..., \lambda_{j,K}^{In} \right).
\label{eq:FixedPoint_Heter}
\end{equation}
The problem of existence and uniqueness of a solution to the non-linear system (\ref{eq:FixedPoint_Heter}) is out of scope of the  present paper. As the performance functions $\mathcal{F}_{j,k}$ may not be explicitly determined in terms of input parameters, the practical determination of a solution to 
(\ref{eq:FixedPoint_Heter}) generally involves a numerical iterative procedure, e.g. a fixed-point algorithm. 

\subsection{Homogeneous network}
\label{Sec:HomogCells}
Now assume that the network consists in \textit{homogeneous cells} in the following sense:
\begin{itemize}
\item[\textit{(i)}] all intrinsic parameters (capacities, arrival rates, ...) are the same for all cells, so that performance functions do not depend on the cell index, that is, $\mathcal{F}_{j,k}(.) = \mathcal{F}_k(.)$; 
\item[\textit{(ii)}] handover routing is homogeneous, i.e., for each class 
$k$, cell $i$ receives handover traffic from a set $\mathcal{J}_k(i)$ of neighboring cells with identical probability $p_k(j,i) = 1 / J_k$, where $J_k$ is the common cardinal of sets $\mathcal{J}_k(i)$. 
\end{itemize}

\noindent
Clearly, any solution to the simpler system
\begin{equation}
\left\{
\begin{array}{ll} 
\lambda_{i,k}^{In} = \lambda_{j,k}^{In} = \lambda_k^{In}, 
\quad \quad \quad \quad \quad \quad \; \; \; \; \forall \, i,j,k,
\\ \\
\lambda_k^{In} = \lambda_k^{Out} = 
\mathcal{F}_k \left( \lambda_{1}^{In}, ..., \lambda_{K}^{In} \right), \; \; \; \, \forall \, k,
\end{array} \right.
\label{eq:FixedPoint_OneCell}
\end{equation}
will provide a particular solution to (\ref{eq:FixedPoint_Heter}), 
hence \textit{the} solution if uniqueness is ensured. For a homogeneous network, the problem thus reduces to the study of a single cell, hereafter called the \textit{representative cell}. System (\ref{eq:FixedPoint_OneCell}) expresses the fact that the outgoing handover traffic is fed back in a balanced way as a supplementary traffic to the ingress of this cell. 

\section{Generic Model}
\label{Sec:MarkovMod}

We consider a single cell whose performance model may be used as the generic tool to provide performance equations (\ref{flow2}) or (\ref{eq:FixedPoint_OneCell}) for a heterogeneous or homogeneous network, respectively (we thus drop the cell index $i$ in all notation). 

\subsection{Markovian Modeling}

The considered cell is supposed to be ``small'', i.e., of limited range so that its transmission capacity $C$ can be assumed spatially constant (this capacity is viewed as an input parameter accounting for radio and interference conditions in the considered cellular network). We handle downlink traffic only and suppose that capacity $C$ is equally  shared among all active users present in its service area (this can be implemented by means of a Round-Robin discipline). Following this fair sharing policy, the system occupancy (number of active transmissions at any time) can then be modeled by a Processor-Sharing (PS) queue \cite{Bon03}. 

Class-$k$ users have i.i.d. transmission requests of data volume 
$\Sigma_k$ with mean $\sigma_k$, hence a service rate $\mu_k = C/\sigma_k$.  
We call $T_k$ the \textit{remaining} sojourn time of a mobile user, i.e., the time duration he physically stays in the cell after the transmission has started. We finally denote by $\theta_k = 1/\mathbb{E}(T_k)$ the mean cell departure rate of class-$k$ users, called class-$k$ \textit{mobility rate}; any class $k$ where $\theta_k = 0$ will be called \textit{static}.

Given this setting, the occupation state of the considered cell can be 
described by the multi-dimensional random process 
$\mathbf{N}(t) = (N_1(t),...,N_K(t))$, where 
$N_k(t)$ is the number of ongoing class-$k$ data transfers in the cell at time $t$. The evolution of this process is represented by a Processor-Sharing queue with impatience \cite{Coff94}, the ``impatient'' customers here corresponding to mobile users that may leave the system before their service completion.

We further assume that $\Sigma_k$ and 
$T_k$ are exponentially distributed with parameters $1/\sigma_k$ and 
$\theta_k$, respectively. Process $\mathbf{N}(t)$ is then Markovian in $\mathbb{N}^K$; from state $\mathbf{n} = (n_1,...,n_K)$ and for 
$\mathbf{e}_k = (0,...,1,...0)$ with 1 corresponding to the $k$-th component, 
we can reach state $\textbf{n} + \textbf{e}_k$ with transition rate 
$\lambda_k$, or state $\mathbf{n} - \mathbf{e}_k$ with transition rate 
$n_k \mu_k / L(\mathbf{n}) + n_k \theta_k$ if $n_k > 0$, 
$L(\mathbf{n}) = \sum_{1 \leq j \leq K} n_j$ denoting the total number of active users.

\begin{prop}
\textbf{\textit{Let $\rho_k = \lambda_k / \mu_k$ be the offered load of class 
$k$. The Markov occupancy process $\mathbf{N}(t)$ has a stationary regime if, and only if,}}
\begin{equation}
\rho_S = \sum_{k \in S}\rho_k < 1,
\label{eq:stabcond}
\end{equation}
\textbf{\textit{where $S$ is the set of static user classes.}}
\label{stability}
\end{prop}
 
See proof in Appendix \ref{SIM}. Note that stability condition 
(\ref{eq:stabcond}) does not depend on the traffic intensity of mobile (i.e., impatient) users; this is intuitively clear since the latter always leave the cell after a finite time and therefore cannot cause a system overload. Now, given (\ref{eq:stabcond}), the equilibrium equations of process $\mathbf{N}(t)$ 
in stationary regime can be written as 

\begin{align}
& \sum_{k=1}^K \left [ \lambda_k + n_k \left ( \frac{\mu_k}{L(\mathbf{n})} + \theta_k \right ) \right ] \Pi (\mathbf{n}) = \sum_{k=1}^K \lambda_k \, \Pi (\mathbf{n} - \mathbf{e}_k) \; +
\nonumber \\
& \sum_{k=1}^K (n_k  + 1) \left ( \frac{\mu_k}{L(\mathbf{n}) + 1} + \theta_k \right ) \Pi (\mathbf{n} + \mathbf{e}_k),
\label{eq:Global_Balance}
\end{align}
with $\sum_{n \in \mathbb{N}^K} \Pi(\mathbf{n}) = 1$, where 
$\Pi(\mathbf{n}) = \mathbb{P}(\mathbf{N} = \mathbf{n})$. 
For given $k$, multiplying each equation of (\ref{eq:Global_Balance}) by $n_k$ and then summing over all state vectors $\mathbf{n} \in \mathbb{N}^K$ provides the following relation.

\begin{lemma}
\textbf{\textit{For any class $k$, the average arrival and departure rates verify the conservation law}}
\begin{equation}
\lambda_k = \mu_k \, \mathbb{E} \left ( \frac{N_k \, \mathbf{1}_{N_k > 0}}{L(\mathbf{N})} \right ) + 
\theta_k \, \mathbb{E}(N_k).
\label{eq:Conserv_Law}
\end{equation}
\end{lemma}

Clearly, process $\mathbf{N}(t)$ is not reversible unless all classes are 
static; its stationary distribution $\Pi(\cdot)$ is thus not amenable to a simple closed form. We can nevertheless evaluate this distribution by solving system (\ref{eq:Global_Balance}) numerically and derive the performance indicators of interest. We first calculate the average throughput $\gamma_k$ of a class-$k$ user. Since mobile users may transfer only a part $X_k$ of their total data volume $\Sigma_k$ during their sojourn in the cell, their average throughput may be evaluated as follows. By Little's law, 
$\mathbb{E}(N_k)/\lambda_k$ is the mean time to transfer the average volume 
$\mathbb{E}(X_k)$ hence
\begin{equation}
\gamma_k = \frac{\lambda_k \mathbb{E}(X_k)}{\mathbb{E}(N_k)}.
\label{eq:Gamma1}
\end{equation}
The average transferred volume $\mathbb{E}(X_k)$ is not directly computable from distribution $\Pi(\cdot)$. We can nevertheless observe that $\lambda_k \mathbb{E}(X_k)$ represents the \textit{carried} traffic intensity for class 
$k$; in a way similar to the derivation of Little's formula, this traffic intensity can be shown to equal the mean bandwidth $\mathbb{E}(\phi_k)$ allocated to class-$k$ users, so that (\ref{eq:Gamma1}) eventually reads
\begin{align}
\gamma_k = \frac{\mathbb{E}(\phi_k)}{\mathbb{E}(N_k)} = \frac{C}{\mathbb{E}(N_k)} \; \mathbb{E} \left (\frac{N_k \, \mathbf{1}_{N_k > 0}}{L(\mathbf{N})} \right )
\label{eq:Gammak}
\end{align}
which involves the distribution $\Pi(\cdot)$ of $\mathbf{N}$ only. We similarly define the \textit{handover probability} for class-$k$ users as the proportion of users that exit the cell before the completion of their transmission, that is,
\begin{equation}
H_k = \frac{\lambda_k^{Out}}{\lambda_k} = \frac{\mathbb{E}(N_k) \, \theta_k}{\lambda_k}.
\label{eq:Hk}
\end{equation}

Using conservation law (\ref{eq:Conserv_Law}) in conjunction with definitions 
(\ref{eq:Gammak}) and (\ref{eq:Hk}) readily yields the following property, which clearly shows the close relationship between the average throughput and the handover probability (except for classes of static customers where 
$\theta_k = 0$ and $H_k= 0$, obviously).

\vspace{0.1in}
\begin{corol}
\textbf{\textit{For any class $k$, the handover probability $H_k$ and the average throughput 
$\gamma_k$ are related by}}
\begin{equation}
H_k = \frac{\theta_k \, \sigma_k}{\gamma_k + \theta_k \, \sigma_k}.
\label{eq:KPI_Relation}
\end{equation}
\label{C1}
\end{corol}
\noindent

In the rest of the paper, we will concentrate on the essential case of two traffic classes, static users (with class index ``$S$'') and mobile users (with 
index ``$M$''). We thus set $\theta_S = 0$, $\theta_M > 0$ and define loads 
$\rho_S = \lambda_S / \mu_S$, $\rho_M = \lambda_M / \mu_M$. 

\subsection{Quasi-Stationary Approximation}
\label{Sec:App_Model}

Although the Markovian model can be numerically solved, it does not lend itself to explicit expressions for performance indicators. 
To circumvent the explosive computation time of both the simulation and the numerical resolution of balance equations, we now develop an approximation framework suitable for the previous two-class scenario. Note 
that an asymptotic for the reneging probability (equivalent to our handover probability) in a multiclass Processor Sharing system with impatience has been provided in \cite{Gui15}, although in a heavy loaded cell with mobile users only.

The underlying idea for the proposed approximation is a Quasi-Stationary 
(QS) assumption. First, consider that the largest the load, the highest tendency of mobile users to leave the cell before the end of their transmission. The stationary distribution of their number might thus be approximated by that of a queuing system without any constraint, i.e., a Poisson distribution describing the occupancy of an M/G/$\infty$ queue. The mean value $A$ of this Poisson distribution remains to be evaluated. 
Now, given a number $m$ of mobile users in the cell, we assume it to remain constant in time (\textit{Quasi-Stationary}) so that the number $N_S$ of static users evolves in a queue with $m$ permanent users. This enables us to obtain the conditional distribution of $N_S$, given $N_M = m$,  thence the joint distribution $\Pi = \Pi^{(1)}$ of the pair $(N_S,N_M)$ and all performance indicators of interest. 

This first step, however, does not provide accurate enough values for these indicators (in particular, $\gamma_M$) when comparing them to that of the 
exact Markovian Model. We thus iterate the QS approximation scheme by fixing now the prior distribution of $N_S$ from the first step estimation 
$\Pi^{(1)}$ obtained earlier; given $N_S = \ell$, we similarly applies the Quasi-Stationary assumption to $N_M$ so as to obtain a second step estimation for the joint distribution 
$\Pi = \Pi^{(2)}$ of $(N_S,N_M)$ and the corresponding evaluation of performance indicators. At this stage, the latter evaluations indeed provide more accurate results (see Sections \ref{Sec/App_ValQS} and \ref{Sec:Discuss}) than that obtained at the first step of the approximation. It has been observed that further iterations of this QS scheme do not bring better accuracy. 

We now specify the QS approximation as follows (see Appendix \ref{QSA} for detailed developments).

\begin{prop}
\textit{\textbf{Let $A > 0$ be the solution to
\begin{equation}
e^{-A} \, (1 - \rho_S) = \frac{\theta_M}{\mu_M} A + 1 - \rho_S - \rho_M.
\label{eq:Fixed_Point}
\end{equation}
Then the throughputs $\gamma_S$ and $\gamma_M$ can be approximated by
\begin{equation}
\left\{
\begin{array}{ll}
\gamma_S = \displaystyle \frac{C (1 - \rho_S)}{1 + A},
\\ \\
\gamma_M = \displaystyle
\sigma_M \left ( \frac{\lambda_M}{\mathbb{E}(N_M)} - \theta_M \right )
\label{eq:Gamma_QS}
\end{array} \right.
\end{equation}
with $\mathbb{E}(N_M) = \sum_{(\ell,m) \in \mathbb{N}^2} m \Psi(m \vert \ell)q(\ell)$, where
\begin{equation}
\left[
\begin{array}{ll}
q(\ell) = e^{-A \rho_S} \, \rho_S^\ell  \displaystyle 
(1 - \rho_S) \sum_{k = 0}^\ell \binom{\ell}{k} \frac{[A (1 - \rho_S)]^k}{k!},
\\ \\
\Psi(m \vert \ell) = \Psi(0 \vert \ell) \displaystyle 
\left ( \frac{\rho_M}{\rho_{\theta}} \right ) ^m \frac{1}{m!} 
\prod_{k=1}^m \frac{\ell + k}{\ell + k + 1 / \rho_{\theta}}
\end{array} \right.
\label{qPsi}
\end{equation}
for all $(\ell,m) \in \mathbb{N}^2$, with factor $\Psi(0 \vert \ell)$ determined by the normalization condition 
$\sum_{m \geq 0} \Psi(m \vert \ell) = 1$ and with constant $\rho_\theta = \theta_M/\mu_M$.}}

\textit{\textbf{The handover $H$ for mobile users is estimated from definition (\ref{eq:Hk}) and the above expression of 
$\mathbb{E}(N_M)$}}.
\vspace{0.1in}
\label{QS_prop}
\end{prop}

\section{Model validation}
\label{Sec:NumRes}

\begin{figure*}[!t]
\begin{center}
\includegraphics[width=5.95cm]{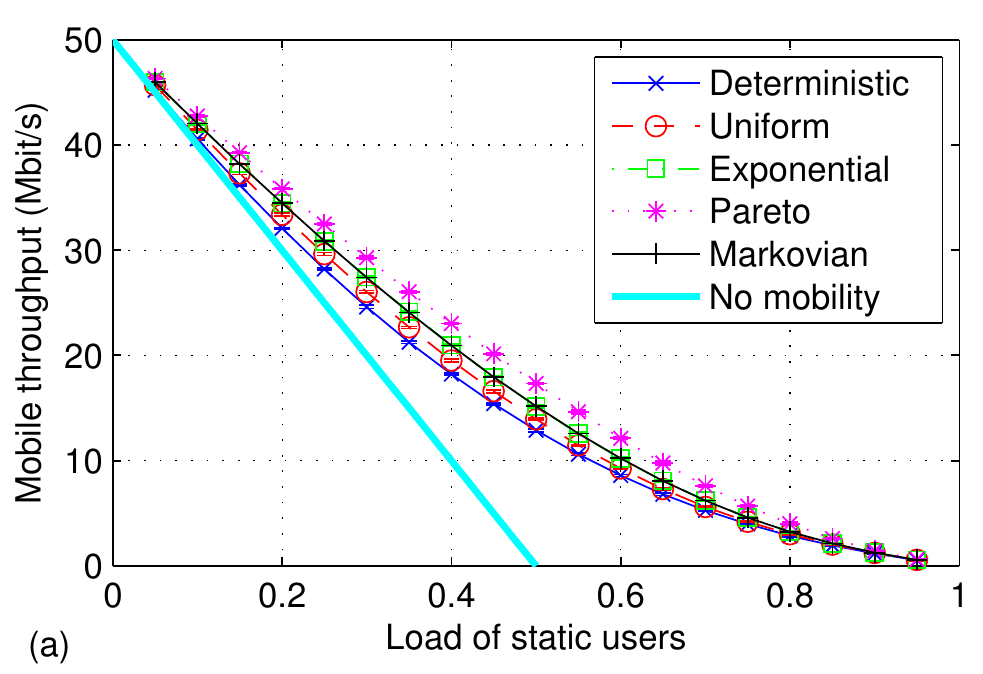} 
\includegraphics[width=5.95cm]{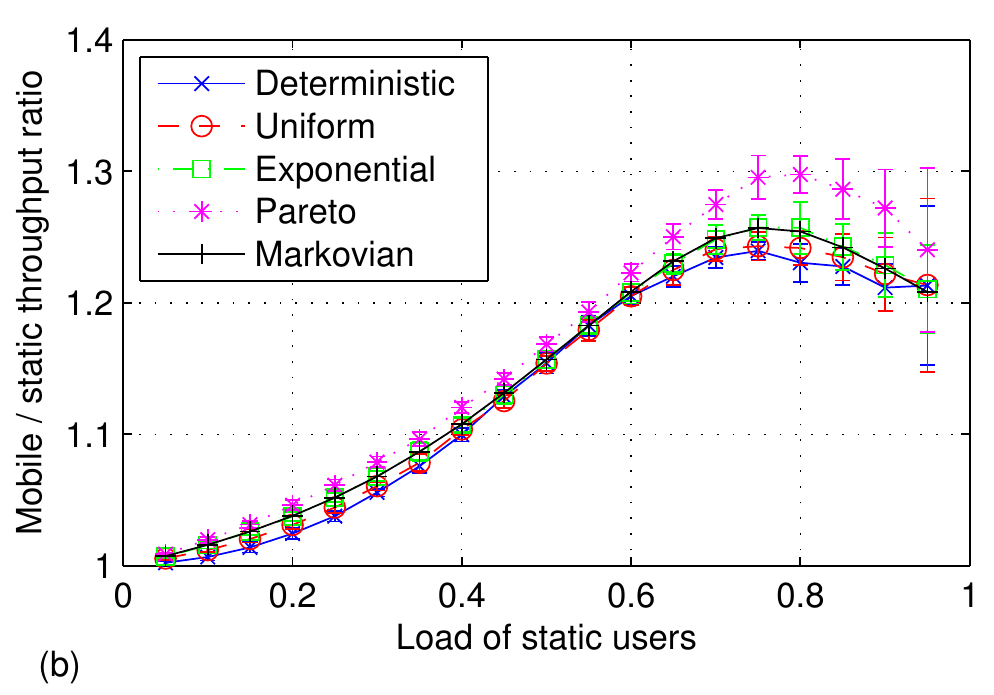} 
\includegraphics[width=5.95cm]{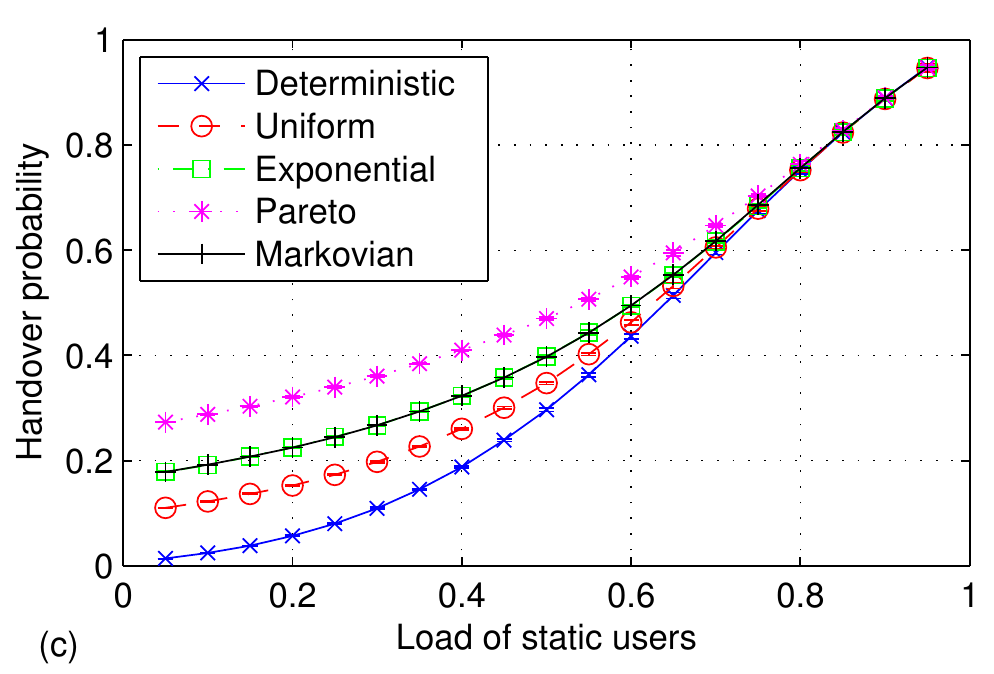}
\caption{Impact of the mobile sojourn time distribution on (a) mobile user throughput, (b) mobile throughput to static throughput ratio and (c) handover probability, computed from simulation and the Markovian IM (proportion of 50\% mobile users and mobility rate $\theta_M = 0.1$ s$^{-1}$).} 
\label{fig:Imp_Deb_Mob50}
\end{center}
\end{figure*}

This section aims at validating the robustness of the proposed approach; to do this, we will distinguish
\begin{itemize}
\item the \textit{impatience model} (IM) consisting in the Markovian model analyzed in Section \ref{Sec:MarkovMod}, where the arrival rate $\lambda_k$ for any user-class $k$ is an exogeneous input parameter;
\item and the \textit{mobility model} (MM), consisting in the same Markovian model but where, as expressed in Section \ref{Sec:FXPT}, 
Equ.(\ref{flow0})-(\ref{eq:FixedPoint_OneCell}), the arrival rate $\lambda_k$ for any class $k$ is the sum of a given exogeneous arrival rate $\lambda_k^0$ 
(the fresh offered traffic) and the handover arrival rate $\lambda_k^{In}$ adjusted so as to balance the outgoing handover rate 
$\lambda_k^{Out}$. 
\end{itemize}

\subsection{Impatience Model}
\label{Sec/Validation}

We here evaluate the IM with a special emphasis on its robustness to the distributions of the sojourn time $T_M$ of mobile users and of the flow volume $\Sigma$.

\subsubsection{Numerical set-up}
we fix a cell capacity $C = 50$ Mbit/s, a  proportion of 50\% mobile users 
and a mean flow volume $\sigma = 12.5$ MB (100 Mbit) for both classes. The mobile users speed is set to $v = 36$ km/h and the mean distance crossed by mobile users, after their data transfer has started, is $\mathbb{E}(D) = 100$ m, hence a mean exit rate $\theta_M = v/\mathbb{E}(D) = 0.1$ s$^{-1}$. 

Event-driven simulations have been performed at flow level. 
The accuracy of results drawn from simulation has been tightly controlled. Specifically, in every configuration, ten independent simulation runs have been performed, generating around 1 million discrete events each, so as to guarantee a confidence interval with range equal to two standard deviations around the mean. In particular, the confidence intervals plotted on 
Fig. \ref{fig:Imp_Deb_Mob50} are very small and cannot be actually distinguished (except at high load for the throughput ratio, a most sensitive indicator). For better readability, such confidence intervals will not be represented anymore in further results. Also, as a check, note that simulation results perfectly match that of the Markovian model in the case of exponential distributions.

\subsubsection{Influence of sojourn time distribution}
we envisage several distributions for $T_M$ so as to obtain a wide range of values for its variance (with given mean). 
Beside the exponential distribution, we thus consider the Deterministic, the Uniform and the Pareto distribution with power index 2.

For such distributions of $T_M$, Fig. \ref{fig:Imp_Deb_Mob50} depicts the variations of $\gamma_M$, $\gamma_M/\gamma_S$ and $H$ with varying load 
$\rho_S$. We observe that the throughput of each class is only marginally impacted by the distribution of $T_M$, indicating that results derived in the Markovian framework remain valid for more realistic sojourn time distributions. The handover $H$ is, however, noticeably more impacted (particularly at low load) and increases with the variance of $T_M$. It also increases with the load, tending to 1 when $\rho_S \rightarrow 1$. 

\subsubsection{Influence of flow volume distribution}
\begin{figure}[!t]
\begin{center}
\includegraphics[width=7cm]{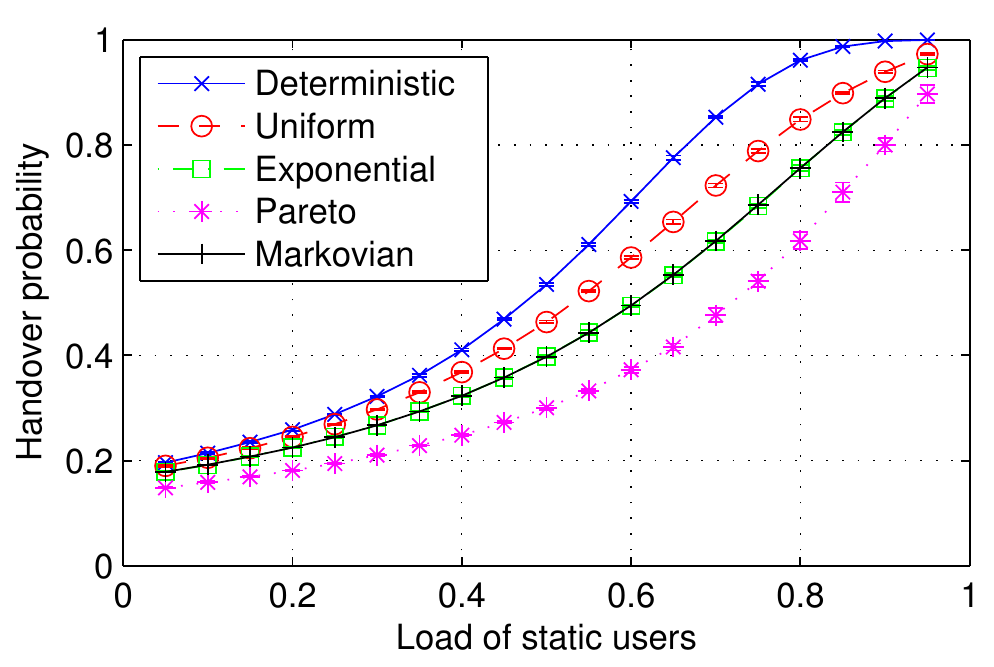}
\caption{Impact of the flow volume distribution on the handover probability, computed from simulation and the Markovian IM for a proportion of 50\% mobile users and a mobility rate $\theta_M = 0.1$ s$^{-1}$.}
\label{fig:Vol_Hdo_Mob50}
\end{center}
\end{figure}

sensitivity tests have also been performed regarding the distribution of the 
flow volume. Surprisingly, the impact of this distribution 
(applied to both mobile and static users) on $\gamma_S$ and $\gamma_M$ is the same as that of Fig. \ref{fig:Imp_Deb_Mob50} with an identical parameter setting (discrepancies between results are indistinguishable and so not shown here); on the other hand, the impact on $H$ is reversed, as shown in Fig. \ref{fig:Vol_Hdo_Mob50}: the greater the distribution variability, the lower $H$.

\subsubsection{Throughput gains}
Fig. \ref{fig:Imp_Deb_Mob50}(a) shows a substantial throughput gain perceived by mobile users compared to a scenario where all users would be static. Note also the different stability regions: $\rho_S < 0.5$ without mobility (with a proportion of 50\% static users) and $\rho_S < 1$ with mobility, as claimed in Proposition \ref{stability}. Fig. \ref{fig:Imp_Deb_Mob50}(b) highlights a significant gain of mobile users throughput over that of static users. 
Such throughput gains result from the opportunistic nature of mobile users who tend to leave the system when it is highly loaded. As indicated by 
Fig. \ref{fig:Imp_Deb_Mob50}(c), the gains are obviously obtained at the expense of larger handover.

\subsubsection{Other scenarios}
we also considered different proportions of mobile users and different values of 
$\theta_M$. Qualitatively, the corresponding performance curves show a behavior quite similar to that of the above scenario. Based on these observations, we will further pay no more attention to the sensitivity of performance indicators and always refer to the Markovian setting with exponential distributions. We have further observed that the throughput gain of mobile over static users appears to be the greatest when the mobility rate is large (say, more than twice the service rate for mobile users) and when the proportion of mobile users is small (say, 20\%).

\subsection{Mobility model in a homogeneous network}
\label{Sec/Multicell}

We now validate our approach for reducing a homogeneous network of small cells to a single representative cell. Handling now the Mobility Model, we stress the fact that the loads (static, mobile or total) as well as the proportion of traffic from each class all refer to the fresh offered traffic, that is, 
$$
\rho_S^0 = \frac{\lambda_S^0 \, \sigma_S}{C}, \quad 
\rho_M^0 = \frac{\lambda_M^0 \, \sigma_M}{C}, \quad 
\rho^0 = \rho_S^0 + \rho_M^0.
$$

\begin{figure}[!t]
\centering
\includegraphics[width=9cm, trim=1cm 18.3cm 0cm 0.05cm, clip]{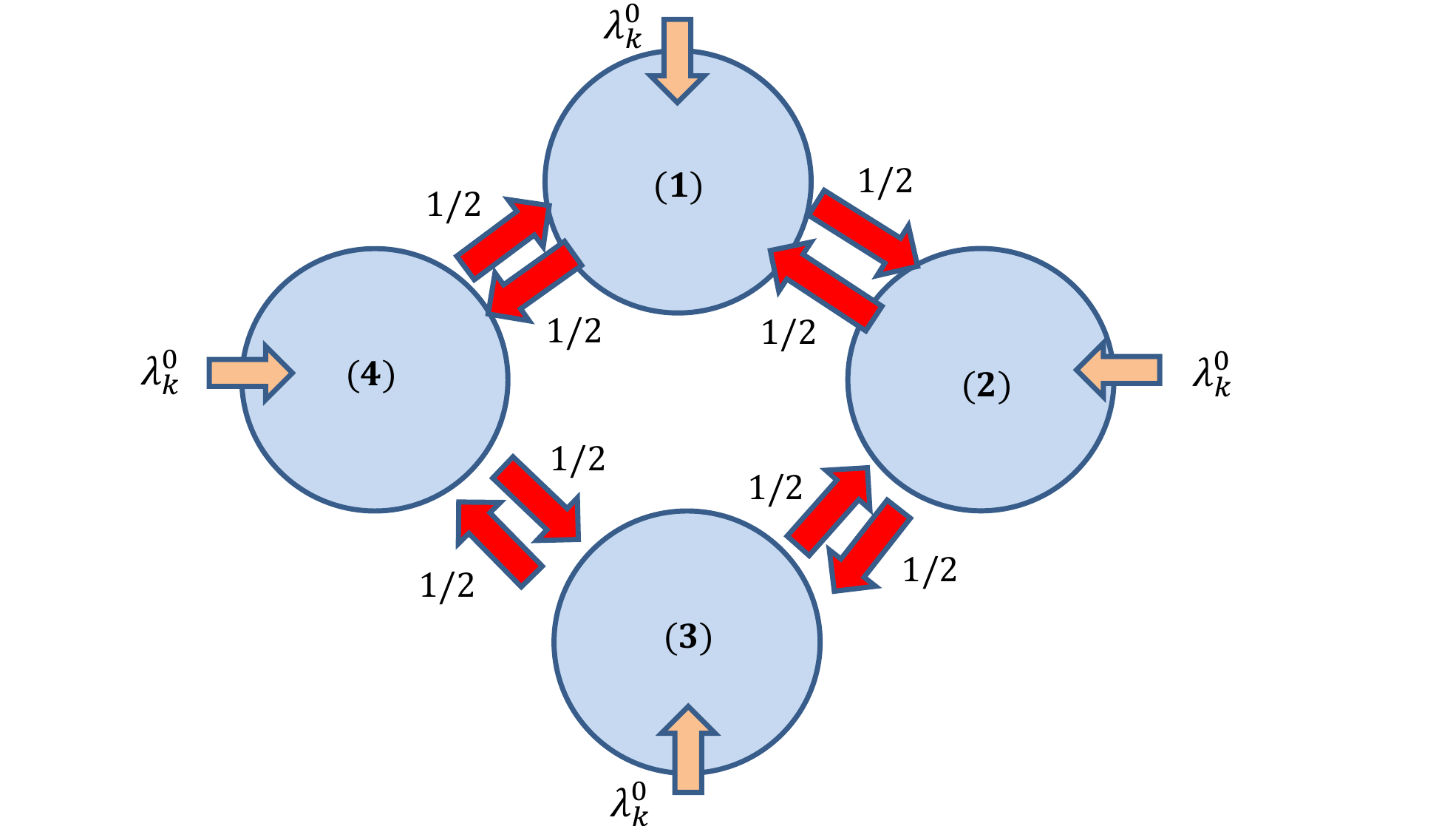}
\caption{A ring network of homogeneous cells with symmetric routing.}
\label{fig:Ring_Network}
\end{figure}

\begin{figure*}[!t]
\begin{center}
\includegraphics[width=5.95cm]{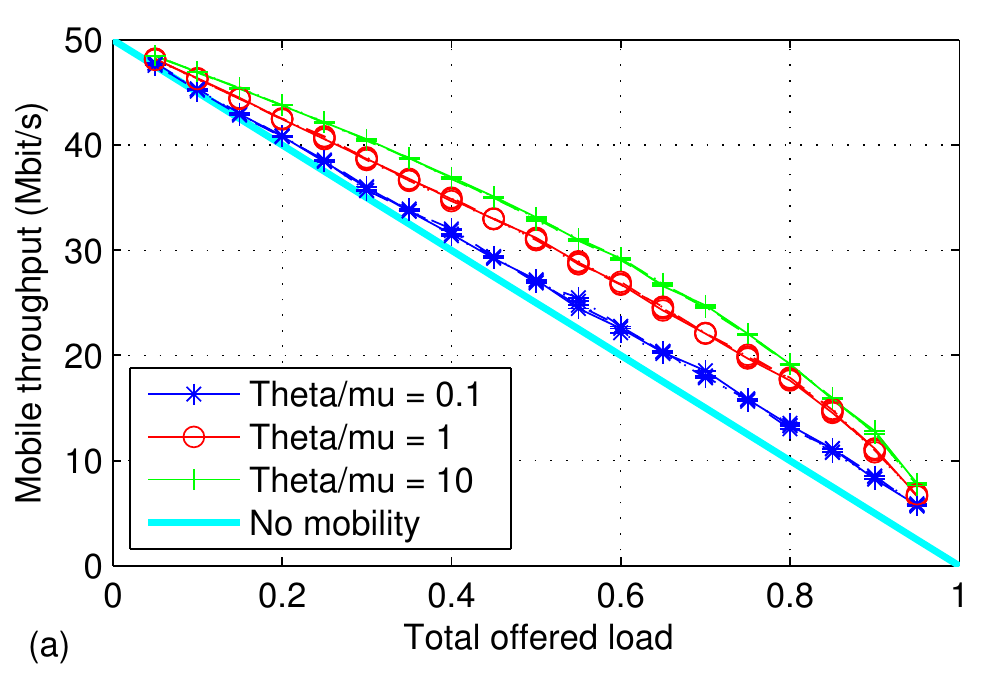}
\includegraphics[width=5.95cm]{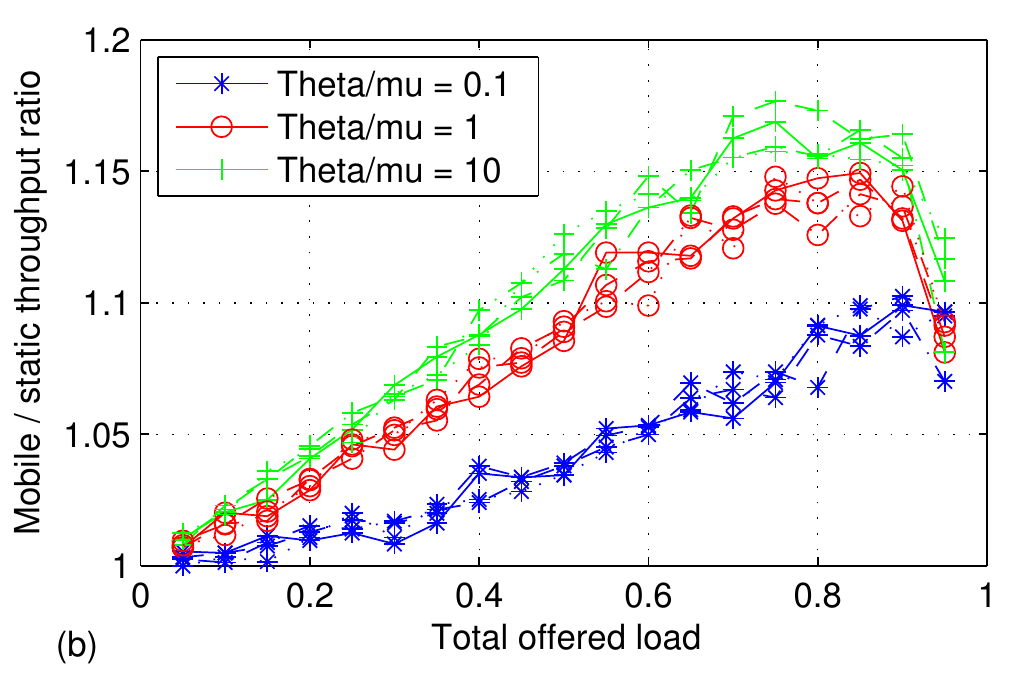}
\includegraphics[width=5.95cm]{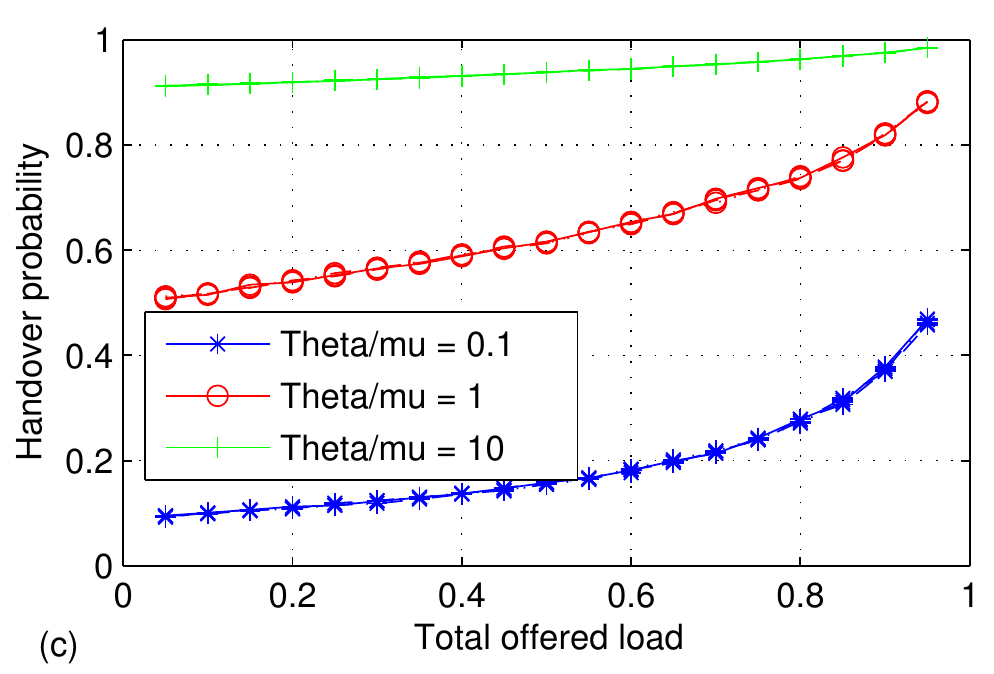}
\caption{Homogeneous ring network: performance indicators, (a) mobile user throughput, (b) mobile throughput to static throughput ratio and  (c) handover probability, obtained from simulation versus the total offered load (initial proportion of 50\% mobile users and $\theta_M / \mu_M =  0.1, 1, 10$).}
\label{fig:Ann_Deb_Mob50}
\end{center}
\end{figure*}

\begin{figure*}[!t]
\begin{center}
\includegraphics[width=5.95cm]{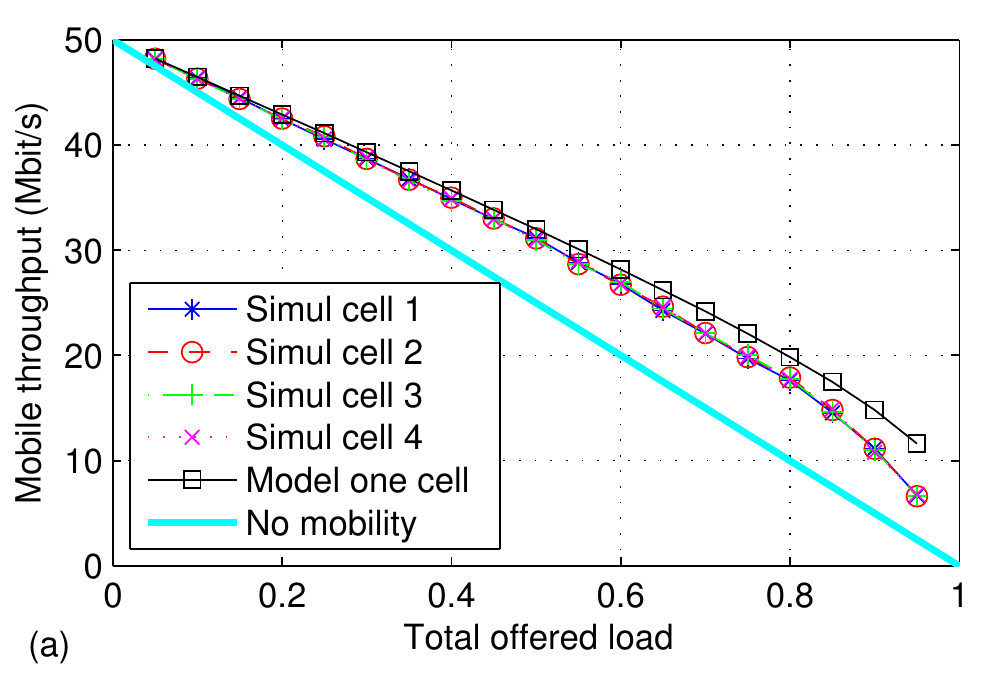}
\includegraphics[width=5.95cm]{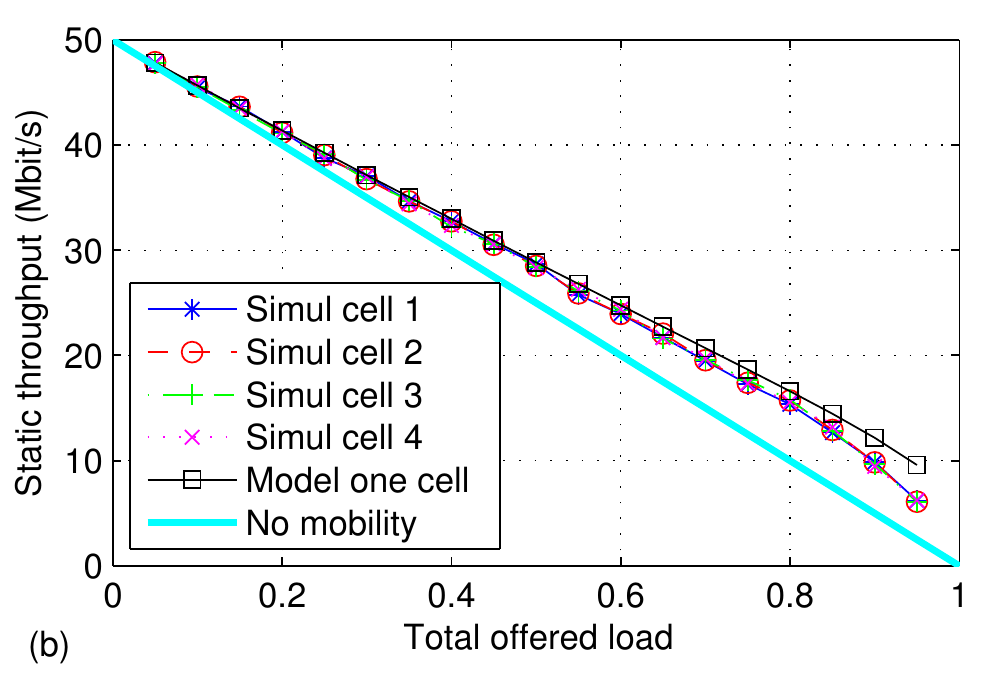}
\includegraphics[width=5.95cm]{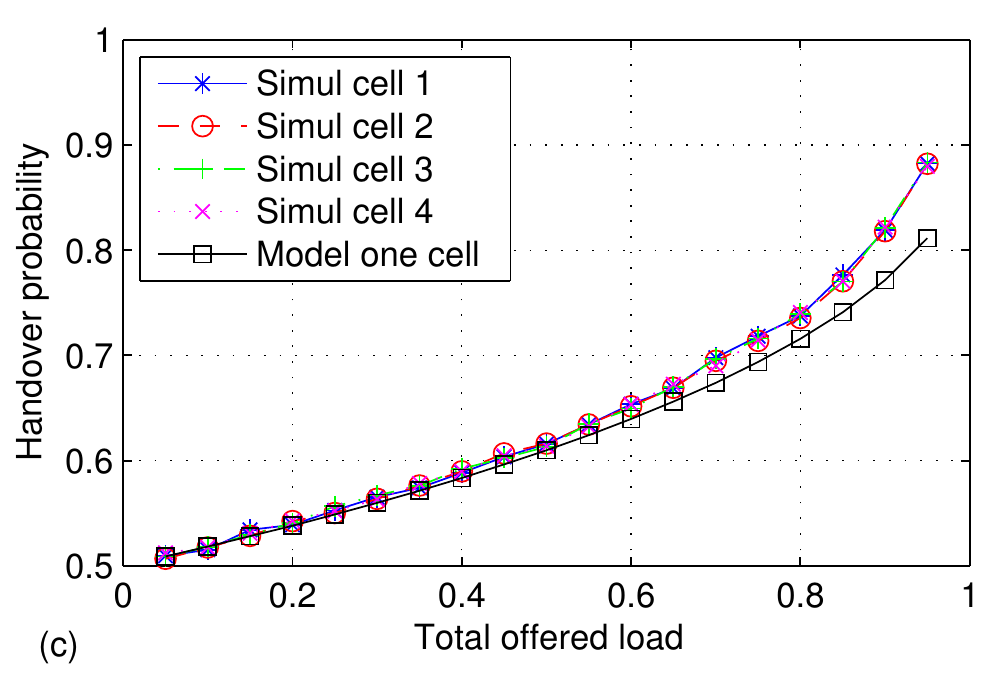}
\caption{Homogeneous ring network: comparison of (a) mobile throughput, (b) static throughput, and (c) mobile handover obtained from simulation with those from the single-cell model (initial proportion of 50\% mobile users and $\theta_M/\mu_M = 1$).}
\label{fig:Imp1_Deb_Mob50}
\end{center}
\end{figure*}

We compare the outputs of simulation experiments on the homogeneous ring network represented in 
Fig. \ref{fig:Ring_Network} to that of the corresponding single-cell mobility model. All four cells of the ring network are equivalent 
and have the same capacity and traffic parameters as in subsection \ref{Sec/Validation}. All random variables are now supposed to be exponentially distributed, and we consider a series of three normalized mobility rates: 
$\theta_M$ equals 0.1, 1 or 10 times the service capacity $\mu_M = 0.5$ $\mathrm{s}^{-1}$.

Fig. \ref{fig:Ann_Deb_Mob50} depicts the performance indicators versus the total offered load (in each cell), obtained from simulation for the three values of the mobility rate. For each value of $\theta_M$, the four curves (each corresponding to a given  cell) are plotted with the same color code; note that they are almost indistinguishable from each other, except for the ratio $\gamma_M/\gamma_S$.

We first observe that the stability region is now characterized by 
$\rho^0 < 1$. This can be easily understood since the considered queuing system is work-conserving, with mobile users re-entering the system until their transfer is completed. 

\vspace{0.05in}
\begin{conject}
\textbf{\textit{In the Mobility Model, condition}} 
$\rho^0 < 1$ \textbf{\textit{ensures the existence of a fixed-point solution to equilibrium equation (\ref{eq:FixedPoint_OneCell}), i.e., 
$\lambda_M^{In} = \lambda_M^{Out}$.}}
\label{ConjectStab}
\end{conject}
\vspace{0.05in}

Note that, for simplicity, this condition will still be called a \textit{stability condition}, as it translates the fact that the resource must have enough capacity to handle all offered traffic. Although the proof of its sufficiency is open, its necessity is straighforward; in fact, by conservation equations 
(\ref{eq:Conserv_Law}) for classes $S$ and $M$ where 
$\theta_M \mathbb{E}(N_M) = \lambda_M^{Out} = \lambda_M^{In}$, we have
$$
\rho_S^0 = \mathbb{E} \left ( \frac{N_S \mathbf{1}_{N_S > 0}}{N_S + N_M} \right ), \, \rho_M^0 = \frac{\lambda_M - \lambda_M^{In}}{\mu_M} = \mathbb{E} \left ( \frac{N_M \mathbf{1}_{N_M > 0}}{N_S + N_M} \right )
$$
hence $\rho^0 = \rho_S^0 + \rho_M^0  < 1$.

From Fig. \ref{fig:Ann_Deb_Mob50}(a)(b), we note that the throughput gains due to mobility increase, quite naturally, with the mobility rate. Other complementary results with various proportions of mobile users confirmed that the mobile/static throughput gain is all the more important that the proportion of mobile users is weak.

The latter simulation results are compared in Fig. \ref{fig:Imp1_Deb_Mob50} to the Markovian single-cell mobility model (\textit{representative cell}) for the case when $\theta_M = \mu_M$. We notice that the single-cell model provides slightly optimistic performance indicators: average throughputs greater and handover probability lower than those from simulation, particularly at high load. From plots obtained for other values of 
$\theta_M/\mu_M$, these discrepancies generally increase with the mobility rate, as expected.

In summary, the presented results provide a number of interesting features, such as the stability region, the good matching between model and simulation, the performance gain of mobile users versus static users, the high sensitivity of handover to various input parameters (mobility rate, proportion of mobile users). The salient point, nevertheless, is the following: there is an inter-cell mobility gain created by the opportunistic displacement of mobile users within the network according to possible local load variations in individual cells. This throughput gain is perceived by both classes (although at a lower level by static users) and is predicted by simulation and by the Markovian single-cell model as well. 

\subsection{Validation of the QS Approximation}
\label{Sec/App_ValQS}

\begin{figure*}[!t]
\begin{center}
\includegraphics[width=5.95cm]{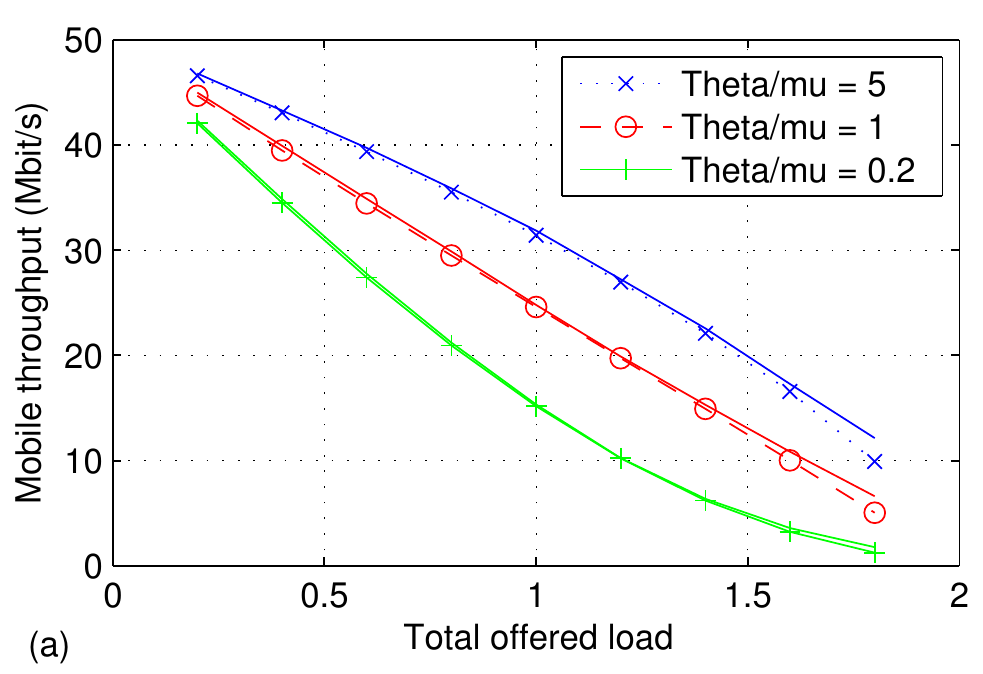}
\includegraphics[width=5.95cm]{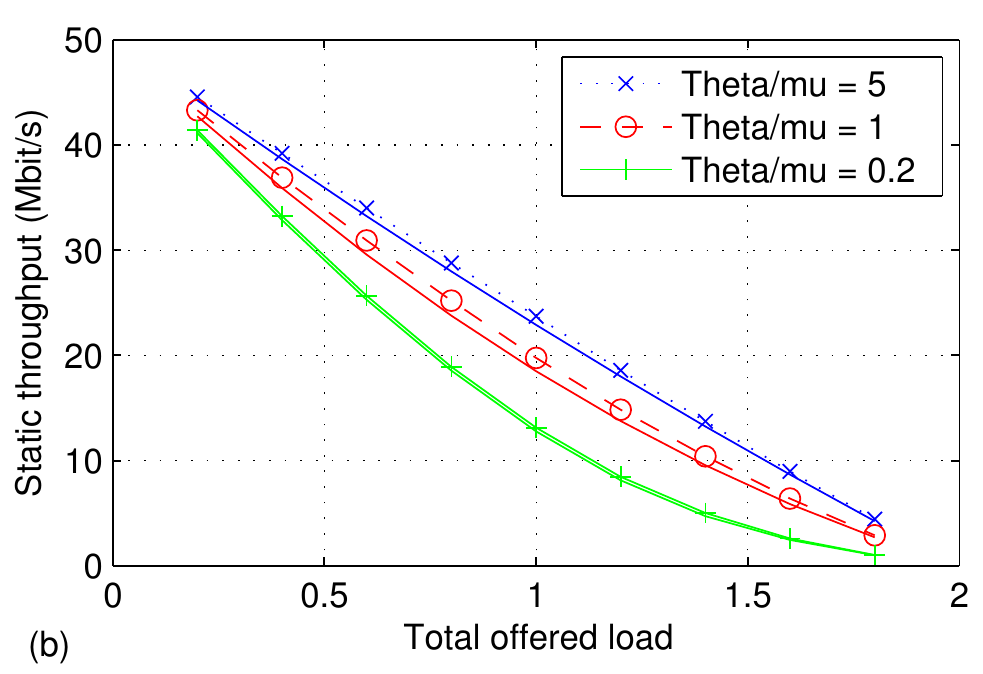}
\includegraphics[width=5.95cm]{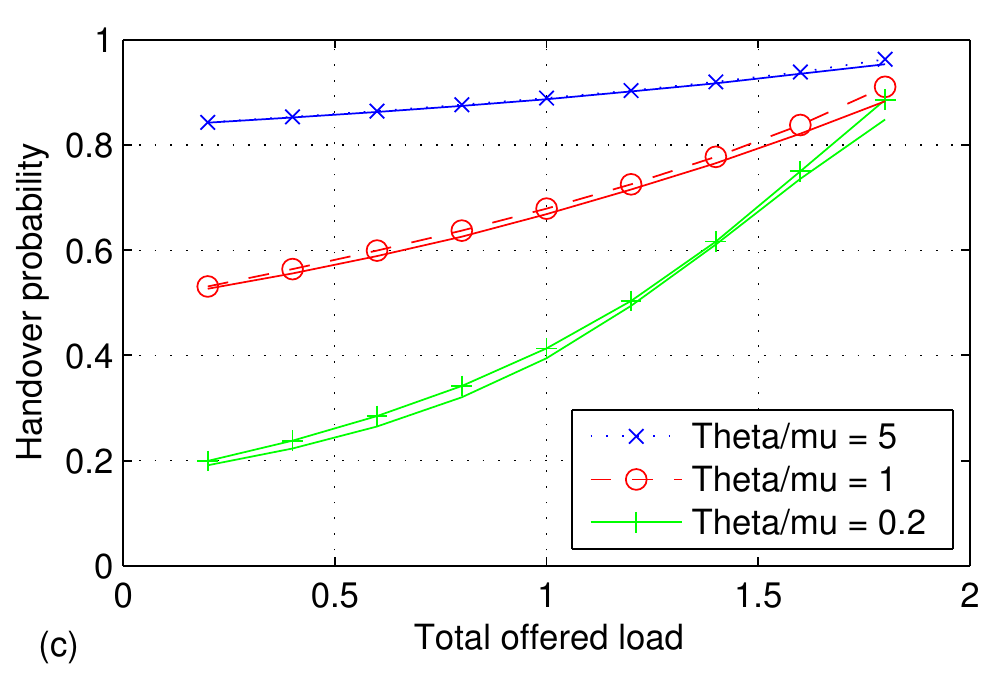}
\caption{Comparison of the (a) mobile throughput, (b) static throughput and (c) handover probability obtained from the Markovian IM (marks) with those from the QS approximation (lines), for a proportion of 50\% mobile users and a normalized mobility rate of 0.2, 1, or 5.}
\label{fig:Aps_Deb_Mob50}
\end{center}
\end{figure*}

The QS approximation is compared in Fig. \ref{fig:Aps_Deb_Mob50} to the results of the Markovian IM, with the same parameter setting as above but with normalized mobility rates of 0.2, 1 and 5. The accuracy of the approximation appears very good for all performance indicators. This has also been checked for other parameter settings, and particularly for other proportions of the mobile user class. 

When comparing the QS approximation to the Markovian MM, however, we observed that the throughput of static users is somewhat lower in the approximation than in the exact model (except when the impact of mobility is very weak, e.g. for a normalized mobility rate = 0.2). We can give the following  interpretation to this: in fact, when applying the first step of the QS approximation to the MM framework, the constant $A$ which solves equation 
(\ref{eq:Fixed_Point}) can be explicitly written as 
$A = \ln(1 - \rho_S) / (1 - \rho^0)$ (after definitions $A = \mathbb{E}(N_M)$, $\lambda_M = \lambda_M^0 + \lambda_M^{In}$ and the balance equation 
$\lambda_M^{In} = \lambda_M^{Out} = \mathbb{E}(N_M) \, \theta_M$ 
between handover arrival and departure rates). Thus, if computed from 
(\ref{eq:Gamma_QS}) after this first step, the static and mobile throughputs should be, as $A$, independent of $\theta_M$. This explains the poor behaviour observed above for the approximate model at this first step. The second step largely improves the accuracy, at least as regards the performance of mobile users, but the throughput of static users remains somewhat less precise and almost constant as can be seen for example in Fig. \ref{fig:Speed_Mob_Mob50} below.

\section{Impact of speed and cell size}
\label{Sec:Discuss}

Interpreting the previous results and figures helps us to assess the impact of the cell size. Indeed, assuming a constant speed $v$, the (residual) mean distance the mobile user travels in the cell 
is $\mathbb{E}(D) = v / \theta_M$; this mean distance typically equals the radius $R$ of a circular cell. Thus for, e.g., $v = $ 90 km/h, the values of $\theta_M$ considered above, namely $5 \, \mu_M$, $\mu_M $ and $0.2 \, \mu_M$ (with $\mu_M = 0.5$ $\mathrm{s}^{-1}$) respectively correspond to a radius of 10 m, 50 m and 250 m, typical of a Femto, Pico, and Micro cell. As expected, users in Femto cells experience the largest throughput since their mobility rate is the highest. As a counterpart, the handover probability they generate is much higher than that in other types of cells. 

\begin{figure*}[!t]
\begin{center}
\includegraphics[width=5.95cm]{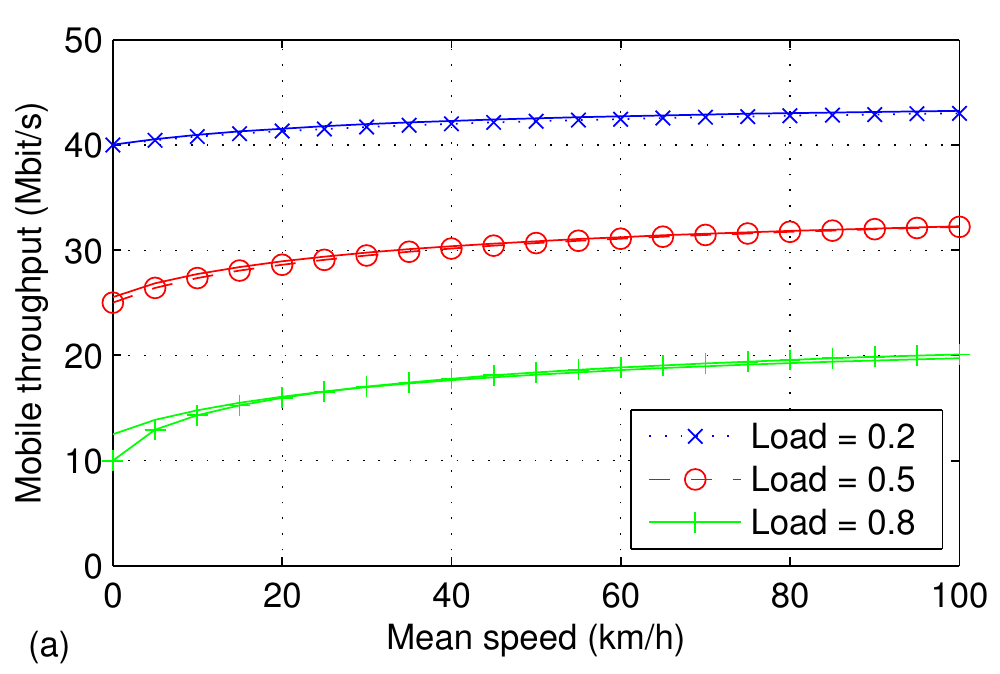}
\includegraphics[width=5.95cm]{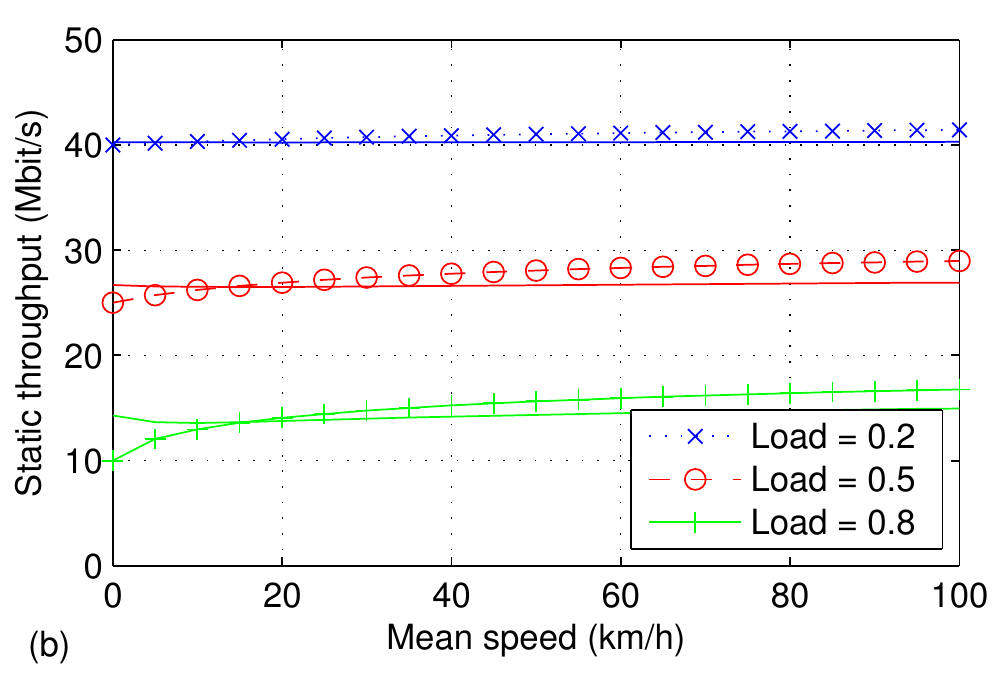}
\includegraphics[width=5.95cm]{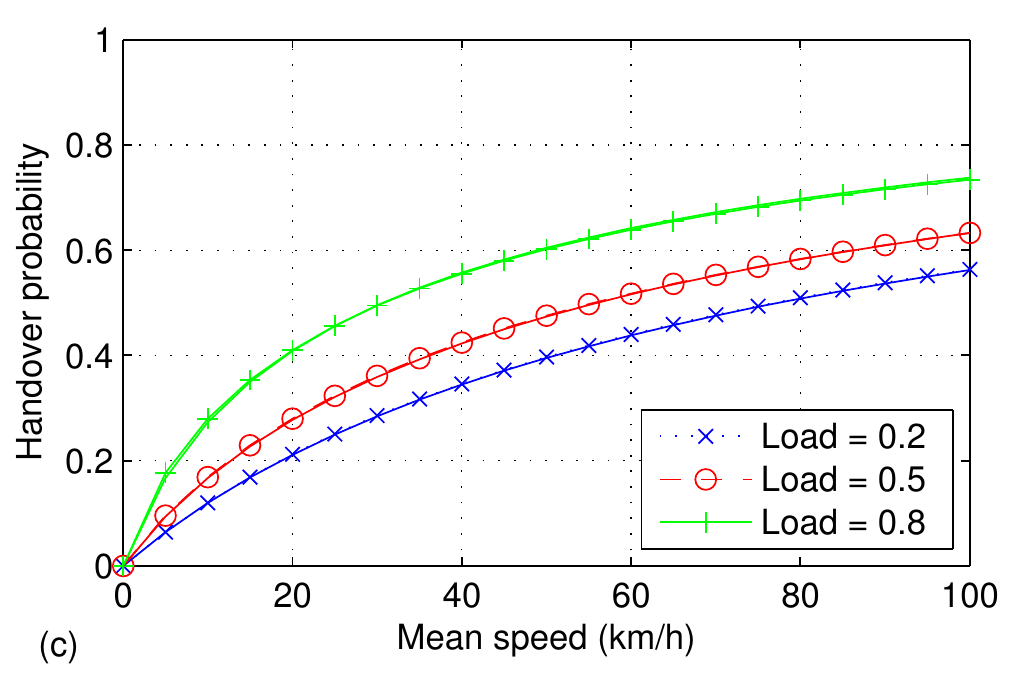}
\caption{Impact of the users speed on (a) mobile throughput, (b) static throughput and (c) handover probability, obtained from the Markovian MM (marked points) and the QS approximation (plain lines), for an initial proportion of 50\% mobile users, cell radius 50 m, and total offered load 0.2, 0.5 or 0.8.}
\label{fig:Speed_Mob_Mob50}
\end{center}
\end{figure*}

Conversely, the values of the mobility rate $\theta_M$ may be interpreted in terms of various mobile speeds for a given cell size. 
Specifically, Fig. \ref{fig:Speed_Mob_Mob50} shows the static and mobile users throughputs and the probability of handover in terms of speed $v$ for different values of the total offered load, 0.2, 0.5, and 0.8, in the framework of the MM. The (initial) proportion of mobile users is still 50\% and the cell radius is 50 m (a Pico cell). In all plots, the results are derived from the exact Markovian model as well as from the QS approximation described above. 

As expected, all performance indicators are increasing functions of the speed. In particular, $H \rightarrow 1$ when $v \rightarrow \infty$, as predicted by 
(\ref{eq:KPI_Relation}); for $v = 0$, both static and mobile users throughputs equal $C (1 - \rho^0)$, the common throughput achieved when all users are static. Again, the QS approximation works nicely for predicting the performance of mobile users, although it is less accurate for the static user throughput. 

\section{Conclusion}

We investigated the impact of inter-cell mobility on user performance in the context of dense networks with small cells. Our  approach relies on two main ideas. First, we have reduced the evaluation of  user performance in a homogeneous network to that of a single representative cell (a generalization to a heterogeneous network has also been introduced). We have then developed simple analytical models that capture mobility through the distribution of the mobile users sojourn time in a cell. A simplifying approximate model has been derived in order to overcome scalability problems of the Markovian model, or in view of possible extensions: based on a Quasi-Stationary assumption, this approximation takes into account the fundamentally distinct behaviors of static and mobile users.

We can notably point out the following outcomes of our study: \textbf{(i)} as a step beyond available studies, we perform the computation of the handover probability, a key performance indicator which mitigates that of the user throughput; 
\textbf{(ii)} our reduction approach to a single representative cell proves reasonable; \textbf{(iii)} throughput performance is robust with respect to different distributions of the mobile users' sojourn time and their flow volume. This insensitivity property justifies the use of Markovian models; 
\textbf{(iv)} the Quasi-Stationary approximation performs quite well in evaluating  performance indicators in various parameter configurations; \textbf{(v)} both classes of users are shown to benefit from a throughput gain induced by the opportunism of mobile users while they travel among cells.

A follow-up to the present study can be envisaged by extending the analysis to cellular networks with spatially varying cell capacities, together with the generalization to heterogeneous cell networks.




\section{Appendix}


\subsection{Stability of the Impatience Model}
\label{SIM}
 
We prove Proposition \ref{stability}. First assume that process 
$(\mathbf{N}(t))_{t \geq 0}$ has a stationary distribution; applying  conservation law (\ref{eq:Conserv_Law}) to each static class $k$ with 
$\theta_k = 0$, then summing over all $k \in S$, gives  
$\rho_S = \sum_{k \in S} \rho_k = \sum_{k \in S} \mathbb{E}(N_k/L(\mathbf{N})) 
< 1$, so that condition (\ref{eq:stabcond}) is necessary. 

Conversely, assume that (\ref{eq:stabcond}) holds, that is, $\rho_S < 1$. For any test function 
$f:\mathbb{N}^K \rightarrow \mathbb{R}^+$, the infinitesimal generator 
$\mathcal{Q}$ of the Markov process $(\mathbf{N}(t))_{t \geq 0}$ is given by
\begin{align}
& \mathcal{Q}f(\mathbf{n}) = \sum_{1 \leq k \leq K} \lambda_k 
\left [ f(\mathbf{n} + \mathbf{e}_k) - f(\mathbf{n}) \right ] \; +
\nonumber \\
& \sum_{1 \leq k \leq K} \left ( \frac{\mu_k n_k}{L(\mathbf{n})} + \theta_k n_k \right ) \mathbf{1}_{n_k > 0} 
\left [ f(\mathbf{n} - \mathbf{e}_k) - f(\mathbf{n}) \right ]
\nonumber 
\end{align}
for $\mathbf{n} \in \mathbb{N}^K$ with 
$L(\mathbf{n}) = \sum_{1 \leq k \leq K} n_k$. By (\cite{Rob03}, Proposition 8.14), $(\mathbf{N}(t))_{t \geq 0}$ is ergodic if there exists a so-called Lyapunov function 
$\Lambda:\mathbb{N}^K \rightarrow \mathbb{R}^+$ and positive constants 
$\eta$, $\delta$ such that
\begin{itemize}
\item[(a)] the set 
$\{\mathbf{n} \in \mathbb{N}^K, \Lambda(\mathbf{n}) \leq \eta \}$ is finite,
\item[(b)] random variables
$\sup_{0 \leq t \leq 1} \Lambda(\mathbf{N}(t))$ and   
$\int_{[0,1]} \vert \mathcal{Q}\Lambda(\mathbf{N}(t)) \vert \mathrm{d}t$ 
are integrable,
\item[(c)] $\mathcal{Q}\Lambda(\mathbf{n}) \leq -\delta$ as soon as 
$\Lambda(\mathbf{n}) > \eta$.
\end{itemize} 

As a candidate Lyapunov function, we consider the function $\Lambda$ defined by
$\Lambda(\mathbf{n}) = \sum_{1 \leq k \leq K} n_k^2/\lambda_k$, 
$\mathbf{n} \in \mathbb{N}^K$. We successively verify conditions (a), (b) and 
(c):
 
$\bullet$ (a) is clearly fulfilled by function $\Lambda$ and any finite $\eta$;

$\bullet$ if $A_k(t)$ is the number of class-$k$ user arrivals within interval $[0,t]$, then $N_k(t) \leq A_k(t) \leq A_k(1)$ for $0 \leq t \leq 1$, where variable $A_k(1)$ has finite first and second moments. The latter inequalities readily ensure the validity of (b) for $\Lambda$; 

$\bullet$ denoting by $M$ the set of mobile user classes and setting 
$\mathbf{x} = \mathbf{n}/L(\mathbf{n})$ for $\mathbf{n} \neq \mathbf{0}$, the above definition of $\mathcal{Q}$ yields
\begin{equation}
\mathcal{Q}\Lambda(\mathbf{n}) = H(\mathbf{x}) + 2 \, F(\mathbf{x}) \cdot L(\mathbf{n}) 
- 2 \, G(\mathbf{x}) \cdot L(\mathbf{n})^2
\label{Gen_Lambda}
\end{equation}
with $H(\mathbf{x}) = K + \sum_{1 \leq k \leq K} x_k / \rho_k$ and
$$
F(\mathbf{x}) = 1 - \sum_{1 \leq k \leq K} \frac{x_k^2}{\rho_k} 
+ \frac{1}{2} \sum_{k \in M} \frac{\theta_k x_k}{\lambda_k}, \; 
G(\mathbf{x}) = \sum_{k \in M} \frac{\theta_k x_k^2}{\lambda_k}.
$$
Note by definition of $L(\mathbf{n})$ that $\sum_{1 \leq k \leq K} x_k = 1$, so that all terms $H(\mathbf{x})$, $F(\mathbf{x})$ and $G(\mathbf{x})$ in 
(\ref{Gen_Lambda}) are bounded. 

For any given $\varepsilon > 0$, it is readily verified that the minimum of 
$\sum_{k \in S} x_k^2 / \rho_k$ under the constraint 
$\sum_{k \in S} x_k \geq 1 - \varepsilon$ is attained at 
$x_k^* = (1 - \varepsilon) \, \rho_k / \rho_S$, $\forall k \in S$, and this minimum equals $(1 - \varepsilon)^2 / \rho_S$. Then,

(i) on the one hand, for $\sum_{k \in S} x_k \geq 1 - \varepsilon$, we have 
$\sum_{k \in M} x_k \leq \varepsilon$ so that the upper bound
\begin{equation}
F(\mathbf{x}) \leq \frac{\rho_S - 1}{\rho_S} + \varepsilon \left ( \frac{2}{\rho_S} + \frac{B}{2} \right )
\label{Freduced}
\end{equation}
holds, with $B = \max_{k \in M} \theta_k / \lambda_k$. Hence, using condition 
$\rho_S < 1$, $F(\mathbf{x})$ can be made smaller than a negative constant 
$- C$ by choosing a small enough $\varepsilon$. It then follows from 
(\ref{Gen_Lambda}) that 
$\mathcal{Q}\Lambda(\mathbf{n}) \leq 
\max_{\mathbf{x}} H(\mathbf{x}) - 2 \, C \cdot L(\mathbf{n})$, 
thus $\mathcal{Q}\Lambda(\mathbf{n})$ tends to $-\infty$ and can be made smaller that any negative constant for large enough $L(\mathbf{n})$;

(ii) on the other hand, for $\sum_{k \in S} x_k < 1 - \varepsilon$ with 
$\varepsilon$ chosen as above, we have $\sum_{k \in M} x_k > \varepsilon$ and thus $G(\mathbf{x}) > A$ with $A = \varepsilon^2  / \vert M \vert \cdot \min_{k \in M} \theta_k / \lambda_k$. If $L(\mathbf{n}) \rightarrow +\infty$ with condition $G(\mathbf{x}) > A$, then $\mathcal{Q}\Lambda(\mathbf{n})$ tends to 
$-\infty$ after (\ref{Gen_Lambda}) and the fact that functions $F(.)$ and 
$H(.)$ are bounded.

We conclude that, in all cases, $\mathcal{Q}\Lambda(\mathbf{n})$ can 
be made smaller than a constant $-\delta < 0$ for large enough 
$L(\mathbf{n})$, or equivalently, for large enough $\Lambda(\mathbf{n})$. 
This fulfills requirement (c).

Conditions (a), (b) and (c) being verified, $\Lambda$ is a Lyapunov function for $(\mathbf{N}(t))_{t \geq 0}$ and condition (\ref{eq:stabcond}) is thus sufficient $\blacksquare$

\subsection{The Quasi-Stationary Approximation}
\label{QSA}
Recall that we now deal with the simpler, but significant, two-class system, $S$ and $M$. As a preamble, applying conservation law (\ref{eq:Conserv_Law}) to each class, we get
$$
\rho_S = \mathbb{E} \left ( \frac{N_S \mathbf{1}_{N_S > 0}}{L(\mathbf{N})} \right ), 
\rho_M = \mathbb{E} \left ( \frac{N_M \mathbf{1}_{N_M > 0}}{L(\mathbf{N})} \right ) + 
\frac{\theta_M}{\mu_M} \mathbb{E}(N_M).
$$
Then, noting that (with $L(\mathbf{N}) = N_S + N_M$)
$$
\mathbb{E} \left ( \frac{N_S \, \mathbf{1}_{N_S > 0}}{L(\mathbf{N})} \right ) + 
\mathbb{E} \left ( \frac{N_M \, \mathbf{1}_{N_M > 0}}{L(\mathbf{N})} \right ) = 
\mathbb{E}(\mathbf{1}_{\mathbf{N} \neq (0,0)}), 
$$
we obtain the relation, which will be useful below:
\begin{equation}
\mathbb{E}(N_M) = \frac{\mu_M}{\theta_M} \left (\rho_S + \rho_M + \Pi (0,0) - 1 \right ).
\label{eq:Conserv_pi0}
\end{equation}

The QS approximation can now be derived in two steps.

\textbf{(I)} First assume that $N_M$ has a Poisson distribution
with mean $A = \mathbb{E}(N_M)$ to be determined. Given $N_M = m$, $N_S$  is supposed to evolve as the occupancy process of a Processor Sharing queue where static users dynamically share the available capacity with a set of $m$ permanent customers. The conditional distribution $\Phi (\cdot \vert m)$ of $N_S$ is derived from the local balance equations of the one-dimensional Markov process $(N_S(t))$, and we obtain
\begin{equation}
\Phi(\ell \vert m) = \Phi(0 \vert m) \, \rho_S^\ell \, \frac{(\ell+m)!}{\ell! \,  m!}, \quad \ell \geq 0,
\label{eq:Extend_Geom}
\end{equation}
with $\Phi(0 \vert m)$ determined by the normalization condition 
$\sum_{\ell \geq 0} \Phi(\ell \vert m) = 1$, hence 
$\Phi(0 \vert m) = (1 - \rho_S) ^{m+1}$. Deconditioning on $N_M$, the joint distribution of $(N_S,N_M)$ then reads
\begin{align}
\Pi^{(1)}(\ell,m) = 
e^{-A} \, \frac{A^m}{m!} \binom{\ell+m}{m} \rho_S^\ell (1 - \rho_S) ^{m+1}
\label{eq:QS2D_Dist}
\end{align}
for $(\ell,m) \in \mathbb{N}^2$ and, in particular, 
$\Pi(0,0) = e^{-A} \, (1 - \rho_S)$. By relation (\ref{eq:Conserv_pi0}) and the latter value of $\Pi(0,0)$, we deduce that the unknown mean $A$ is determined as the unique positive solution to the implicit equation (\ref{eq:Fixed_Point}). 

Applying relations (\ref{eq:Conserv_Law}) and (\ref{eq:Gammak}) with $\mathbb{E}(N_M) = A$  first provides the expression (\ref{eq:Gamma_QS}) for $\gamma_M$. 
As to static users, we use (\ref{eq:QS2D_Dist}) to derive
$\mathbb{E}(N_S) = \rho_S \mathbb{E}(N_S) + \rho_S \mathbb{E}(N_M + 1)$,
hence $\mathbb{E}(N_S) = (1 + A) \rho_S/(1 - \rho_S)$ so that 
the defining relation $\gamma_S = \lambda_S \, \sigma_S/\mathbb{E}(N_S)$ eventually leads to formula (\ref{eq:Gamma_QS}) for $\gamma_S$. 

\textbf{(II)} As argued in the main text, we need to go one step further by considering now that mobile users ``see" a succession of stationary regimes conditioned on the number of static users. First 
compute the marginal distribution $q(\cdot)$ of $N_S$ from the joint  distribution (\ref{eq:QS2D_Dist}) as
$$
q(\ell) = \sum_{m \geq 0} e^{-A} \, \rho_S^\ell  
(1 - \rho_S) \binom{\ell + m}{m} \frac{[A (1 - \rho_S)]^m}{m!}
$$
for all $\ell \geq 0$; the latter formula reduces to expression 
(\ref{qPsi}) after simple manipulations. For given $N_S = \ell$, the number of mobile users is then supposed to evolve as the occupancy process of a Processor Sharing queue with impatient (mobile) customers and a set of $\ell$ permanent customers. The conditional distribution $\Psi(\cdot \vert \ell)$ for the number of mobile users then verifies the local balance equations
\begin{equation}
\lambda_M \Psi(m - 1 \vert \ell) = m \left (\frac{\mu_M}{\ell + m} + \theta_M \right ) \Psi(m \vert \ell)
\label{eq:Local_BalImp}
\end{equation}
for all $m > 0$. By recursion, we then obtain expression (\ref{qPsi}) for 
$\Psi(m \vert \ell)$, $m \geq 0$, with $\Psi(0 \vert \ell)$ given by the normalization condition. By deconditioning on $N_S$, the joint distribution of $(N_S,N_M)$ is now given by
\begin{equation}
\Pi^{(2)}(\ell,m) = 
\Psi(m \vert \ell) \, q(\ell), \quad (\ell,m) \in \mathbb{N}^2.
\label{eq:QS2D_DistBIS}
\end{equation}

The average throughput $\gamma_S$ in this second step QS approximation is unchanged compared to that given by (\ref{eq:Gamma_QS}), since 
$\gamma_S = C \rho_S / \mathbb{E}(N_S)$ and the marginal distribution 
$q(\cdot)$ of $N_S$ is  now known \textit{a priori}. Besides, relations  
(\ref{eq:Conserv_Law}) and (\ref{eq:Gammak}) again enable us to write $\gamma_M$ as in (\ref{eq:Gamma_QS}) where the mean $\mathbb{E}(N_M)$ is now re-evaluated from the joint distribution (\ref{eq:QS2D_DistBIS}). According to definition (\ref{eq:Hk}), the handover probability is similarly re-evaluated 
from $H = \mathbb{E}(N_M) \, \theta_M / \lambda_M$ $\blacksquare$


\end{document}